\RequirePackage{fix-cm}
\documentclass[twocolumn,epjc3]{svjour3}  
\smartqed  % flush right qed marks, e.g. at end of proof
\RequirePackage{graphicx}

\usepackage{amssymb}
\usepackage{amsmath}
\journalname{}
\begin{document}

\title{$F(R,w) Gravity$%\thanksref{t1}
}
\subtitle{A new gravity framework}


\author{Mahmoud AlHallak \thanksref{e1,e2,addr1,addr2}}

\thankstext{e1}{e-mail: mahmoud.alhallak@damascusuniversity.edu.sy}
\thankstext{e2}{e-mail: : phy.halak@hotmail.com}


\institute{Al Ain University \label{addr1}
           \and
            Physics Dept., Damascus University, Damascus, Syria \label{addr2}}

\date{Received: date / Accepted: date}

\maketitle

\begin{abstract}
In this work we present a new framework of the gravity sector by considering the extension $F(R,w)$, in which $R$ is the Ricci scalar and $w$ is the equation of state. Three different choices of function $F(R,w)$ are investigated under the Palatini formalism. The models appear equivalent to $F(R)$ models of gravity with effective momentum-energy tensors. For linear dependence of Ricci scalar in which $F(R,w)=k(w)R$, the model appears equivalent to Einstein-Hilbert action with effective momentum-energy tensor. Recovering the minimal coupling case of the last choice does not face Jordan-Einstein frame ambiguities and exhibits natural alignments with general relativity results in the matter\text{/} radiation dominated eras. We discuss some astrophysical implications of the model by considering scalar fields as dominant matter forms. We show that the Higgs inflation could be saved within the $F(R,w)$ model. We suggest some future investigations exemplified by constant-roll inflation and universe evolution for $F(R)=f(R)k(w)$ where $f(R)$ represents the Starobinsky gravitational form.  Using the model and comparing it with pure $F(R)$ gravity, we provide preliminary indications of $F(R,w)$'s impact. As a final note, we suggest using the Polytropic equation of state in future works to investigate $F(R,w)$.
\end{abstract}

\section{Introduction:}
The theory of general relativity GR \cite{Einstein:1916vd} has made significant contributions to cosmology, which cannot be ignored. The theory of general relativity alters our understanding of space-time and its relationship to energy in some way that we cannot avoid in any future developments of our knowledge of the universe. 

In spite of the previous notes, GR, under its original minimal metric formalism, failed to explain some recent cosmological observations, such as,  the late expansion of the universe measured from the light-curves of several hundred type Ia supernovae \cite{SupernovaSearchTeam:1998fmf,SupernovaCosmologyProject:1998vns}. \\
Several interesting attempts are made in the scientific community to modify GR in order to solve and explain these cosmological puzzles. 
\\
$F(R)$ gravity \cite{Carroll:2003wy,Hu:2007nk} is the most common modification, in which the gravitational sector is represented by higher-order curvature. Many cosmological features are examined within F(R) gravity, such as in \cite{Liu:2023ypa,Odintsov:2023nqe,Odintsov:2023cli,AlHallak:2022gbv,AlHallak:2021hwb,AlHallak:2023wsx}. 
\\One of the interesting extensions of $F(R)$ models is assuming a non-minimal coupling between the curvature represented by Ricci scalar and matter represented by the momentum-energy tensor $T$. This class of models is known as $F(R,T)$  gravity \cite{Harko:2011kv}. 
\\
Models of this category are investigated both from a metric perspective \cite{Harko:2011kv} as well as a Palatini perspective\cite{Wu:2018idg}. 
\\
A wide range of cosmological aspects have been studied in $F(R,T)$ gravity. In this section, we will mention some of these studies that can serve as motivation for performing similar studies in the new model of $F(R,w)$ gravity. 

A study of wormholes that involves electric charge effects is reported in \cite{Azmat:2023ygn}. Authors use $F(R, T)$ gravity to investigate the existence and viability of wormholes. A wormhole shape function is obtained, the fundamental wormhole conditions are analyzed, and the validity of the null energy condition is examined. In addition, the authors analyze the gravitational active mass of wormhole solutions based on their electric charge dynamics. These results demonstrate that the obtained wormhole solutions exhibit valid behavior and that the values of the complexity factor are always within reasonable limits.
\\
Other studies pertaining to wormholes in $F(R,T)$ can be found in \cite{Moraes:2019pao,Moraes:2017rrv,Moraes:2016akv,Sahoo:2017ual,Bhar:2021lat}. 

A study of cosmological inflation is carried out in F(R,T) gravity. Ref \cite{Gamonal:2020itt} investigates slow-roll inflation using an explicit form of $F(R,T)=R+ 2 \kappa \alpha T$. Authors analyze observable quantities represented by the spectral index $n_s$ and tensor-to-scalar ratio $r$. They computed these quantities for a general power-law potential, natural and quasi-quartic Hilltop inflation, and a Starobinsky model.
\\ 
The ref \cite{Deb:2022hna} investigates the Double-Well potential using the same function of $F(R,T)$ and analyzes $n_s$ and $r$ as the primary observables. 
\\
In \cite{Bhattacharjee:2022lcs} authors employ mimetic $F(R,T)$ gravity coupled with Lagrange multiplier and mimetic potential to yield viable inflationary cosmological solutions consistent with latest Planck and BICEP2/Keck Array data.

As in \cite{Singh:2020gxd} and \cite{Sun:2015yga}, gravity extension $F(R,T)$ has been explored as a possible explanation for late accelerating expansion of the universe. 
\\
The dark energy models are also investigated in $F(R,T)$ such as in \cite{Maurya:2022pzw} where a perfect dark fluid is examined in  Logarithmic $F(R,T)$ Cosmology with $F(R,T)=R-\kappa \alpha \ln(T)$.
\\
Holographic dark energy has been the subject of several studies, as can be seen in \cite{Sharma:2022dzc,Shaikh:2022ynt,Varshney:2021xvg,Shaikh:2019ppk}.

A second interesting study in this field is the $F(R,L_m)$ gravity \cite{Harko:2010mv}.
In a similar manner to $F(R,T)$ gravity, various cosmological phenomena are investigated in $F(R,L)$. 
Studies on wormholes can be found in \cite{Naseer:2023rab,Kavya:2023tjf,Kavya:2023mwv}. Investigations into dark energy in the context of$ F(R,L_m)$ are discussed in \cite{Maurya:2023lxt,Pradhan:2022msm}.

A recent proposal by Dioguardi et al. \cite{Dioguardi:2023jwa} introduces a model of $F(R,X)$ to examine the inflationary expansion of the universe, where $X$ represents the kinetic energy of the inflaton field.

In the current work, we introduce a novel gravity framework by considering an extended form of gravity denoted $F(R,w)$, where $w$ represents the equation of state. The model exhibits a notable characteristics: 
The model coincides  naturally with General Relativity during epochs in which matter and radiation are dominant. 
A closer look at specific forms of $F(R,w)$ shows changes in the scales of some cosmological quantities such as the Hubble parameter and the cosmological acceleration of the universe, but the functions of these quantities remain unchanged. This statement, however, needs further investigation before it can be generalized.
The model also has a significant impact when scalar fields become dominant during early stages of the evolution of the universe (inflaton fields) or during the later expansion phase (quintessence). 
Unlike common models of non-minimal coupling, for $F(R,w)=R+\alpha R w$ under Palatini formalism, the model appears to be free of Jordan-Einstein frame ambiguity since it is reduced to the minimal case without the need for conformal transformation.

The paper is organized as follows. In section. 2, $F(R,w)$ is presented in its most general form, analyzing it both in the metric and Palatini formalisms.
Section. 3 discusses specific cases of model $F(R,w)$ and derivations of field equations and Friedmann equations based on FRW flat metric. Further, we discuss the reduction of the considered models to the minimal coupling case. Finally, In section. 4, the astronomical implications of the model are illustrated through an examination of the inflaton field as dominating matter during early universe evolution. We consider the Higgs inflationary potential, and we show how it can be saved within $F(R,w)$ gravity. As a follow-up, we provide some comments for future studies exemplified by constant-roll inflation, the Starobinsky model, and polytropic equations of state, and we provide some insights into how $F(R,w)$ impacts them. We end up with a summary and conclusion in section. 5.
\section{The model:}
As a first step, we will discuss the most general case of $F(R,w)$ gravity, in which the action is expressed as follows: 
\begin{equation}\label{eq:1}
S=\frac{1}{16\pi G} \int F(\mathcal{R} ,w) \sqrt{-g} d^4x + \int \mathcal L_M \sqrt{-g} d^4 x 
\end{equation}
Throughout this section, we discuss the the field equations in both cases of metric and Palatini formalism, where in the case of metric formalism, the connection is constructed from the metric tensor through the Levi-Civita connection $\Gamma^\lambda_{\mu\nu}=\{^\lambda \: _{\mu\nu}\}$, while in the Palatini formalism, the connection and the metric are considered independent.

According to the metric formalism, by taking the variation of the action \eqref{eq:1} with respect to the metric tensor, the modified field equations take the following form: 
\begin{equation}\label{eq:2}
\begin{split}
F_\mathcal{R}(\mathcal{R},w) \mathcal{R}_{\mu\nu} - \frac{1}{2}F(\mathcal{R},w)g_{\mu\nu} + (g_{\mu\nu}\Box - \nabla_\mu \nabla_\nu)\\ F_\mathcal{R}(\mathcal{R},w)  =8 \pi G T_{\mu\nu}
\end{split}
\end{equation}
By contracting equation \eqref{eq:2}, the relation between the Ricci scalar $\mathcal{R}$ and the trace $T$ of the stress-energy tensor is as follows:
\begin{equation}\label{eq:3}
F_\mathcal{R}(\mathcal{R},w) \mathcal{R} + 3 \Box F_\mathcal{R}(\mathcal{R},w)-2F(\mathcal{R},w)=8\pi G T
\end{equation}
By eliminating the term $\Box F_\mathcal{R}(\mathcal{R},w)$between equations
\eqref{eq:2} and \eqref{eq:3}, the gravitational field equations can be written in the form: 
\begin{equation}\label{eq:4}
\begin{split}
F_\mathcal{R}(\mathcal{R},w) (\mathcal{R}_{\mu\nu}-\frac{1}{3} \mathcal{R} g_{\mu\nu}) + \frac{1}{6}F(\mathcal{R},w)g_{\mu\nu}= \\ 8\pi G (T_{\mu\nu}-\frac{1}{3} T g_{\mu\nu})
\end{split}
\end{equation}

In the Palatini formalism, the gravitational action \eqref{eq:1} is now varied with respect to the metric tensor $g_{\mu\nu}$ under the assumption $\delta R_{\mu\nu}=0$. 
\\As a result we obtain the field equations: 
\begin{equation}\label{eq:5}
F_R(R,w) R_{\mu\nu} - \frac{1}{2}F(R,w)g_{\mu\nu} =8 \pi G T_{\mu\nu}
\end{equation}
The expression for the Ricci scalar is obtained by contracting the equation \eqref{eq:5} with $g_{\mu\nu}$ as, 
\begin{equation}\label{eq:6}
R F_R(R,w) + 2 F(R,w)=8 \pi G T
\end{equation}
In terms of the Einstein tensor one can write: 
\begin{equation}\label{eq:7}
\begin{split}
G_{\mu\nu}=R_{\mu\nu}-\frac{1}{2}g_{\mu\nu}R=\frac{1}{F_R(R,w)} \\ \bigg\{8 \pi G\ (T_{\mu\nu} - \frac{g_{\mu\nu}}{2}T)- \frac{g_{\mu\nu}}{2}F(R,w)\bigg\}
\end{split}
\end{equation}
This paper focuses on palatini formalism due to its interesting characteristics in $F(R,w)$ models, whereas future studies will examine the model in metric formalism.
\section{Particular Cases:}
This section discusses some possible choices regarding the function $F(R,w)$. There are many more forms that could be assumed and studied, but we only provide three examples here. 
\subsection{$F(R,w)= f(R)k(w)$}
Let's consider the following particular function of $F(R,w)$ gravity as: 
\begin{equation}
F(R,w)= f(R)k(w)
\end{equation}
Taking the variation of action \eqref{eq:1} with respect to the metric $g_{\mu\nu}$, we can express the field equation as follows: 
\begin{equation}\label{eq:erh}
f_R R_{\mu\nu}-\frac{1}{2}f(R) g_{\mu\nu}=8\pi G T^{eff}_{\mu\nu}
\end{equation}
Where: 
\begin{equation}
T^{eff}_{\mu\nu}=\frac{T_{\mu\nu}}{k(w)}
\end{equation}
Interestingly, the model can be reduced to an $f(R)$ model with an effective momentum-energy tensor. 
\\
Taking the variation of the action with respect to connection we can find: 
\begin{equation}\label{eq:der_Gamma}
\nabla_\alpha (\sqrt{-g}f_R g^{\mu\nu})=0
\end{equation}
In order to solve equation \eqref{eq:der_Gamma}, it is convinent to introduce a new metric: 
\begin{equation}\label{eq:new_metric}
\sqrt{h}h_{\mu\nu}=\sqrt{-g} f_R g_{\mu\nu}
\end{equation}
The connection for this metric is a Levi-Civita connection. so both metrics are related by the conformal factor. 
\begin{equation}\label{eq:New_metric}
h_{\mu\nu}=f_R g_{\mu\nu}
\end{equation}
For this reason one should assume that the conformal factor $f_R \neq 0$.
\\ 
Considering perfect fluid  
$T_\mu^\nu=diag(-\rho,p,p,p)$
we can find the effective equation of state equivalent to the standard one 
$
w(\text{eff})=\frac{p_{\text{eff}}}{\rho_{\text{eff}}}=w=\frac{p}{\rho}
$
where, $\rho_{\text{eff}}=\rho/K(w)$ and $p_{\text{eff}}=p/K(w)$.

As long as we are in the Jordan frame, we can continue working on the field equations \eqref{eq:erh}. Alternatively, equations can also be displayed in the Einstein frame, where gravitational action is expressed formally in Hilbert-Einstein form.
\\ In the Jordan frame, action \eqref{eq:1} is dynamically equivalent to the first order Palatini gravitational action:
\begin{equation}\label{eq:euy}
\begin{split}
S=\frac{1}{16\pi G} \int  d^4x \sqrt{-g} \big[\frac{1}{2}(f(\zeta)+f^\prime(\zeta)(R-\zeta))\big]+ \\ \int d^4x \sqrt{-g} L_M^{eff}
\end{split}
\end{equation}
with $f^{\prime\prime}(R)\neq 0$.
\\
The action \eqref{eq:euy} takes the Brans-Dicke model form by introducing an auxiliary scalar field $\Phi=f^\prime(\zeta)$ and taking into consideration the constraint $\zeta=R$.
\begin{equation}\label{eq:muy}
S=\frac{1}{16\pi G} \int  d^4x \sqrt{-g} \big[\frac{1}{2}(\Phi R-U(\Phi)\big]+\int d^4x \sqrt{-g} \mathcal L_M^{eff}
\end{equation}
where $U(\Phi)=\zeta(\Phi)\Phi-f(\zeta(\Phi))$.
\\
The action \eqref{eq:muy} can be reduced to the minimal coupling case by applying a conformal transformation $g_{\mu\nu}^E=\Phi g_{\mu\nu}^J $: 
\begin{equation}
S=\frac{1}{16\pi G} \int  d^4x \sqrt{-g_E} (R_E-U_E(\Phi))+\int d^4x \sqrt{-g_E} \mathcal L_{M(E)}^{eff}
\end{equation}
Where $U_E=U/\Phi^2$.
\\
In FRW metric, Friedmann equations take the form:  
\begin{equation}
 H_E^2=\frac{8 \pi G}{3} \bigg(\rho^E_\Phi+\rho^E_M\bigg)
\end{equation}
\begin{equation}
(\dot{H}+H^2)_E=-\frac{4\pi G}{3} \big(2\rho^E_\Phi-\rho^E_M+3P^E_M\big)
\end{equation}
Subscript E denotes the evaluation of all quantities in the Einstein frame, with $\rho^E_M=\rho_M/\Phi k(w)$ and $P^E_M=P_M/\Phi k(w)$. 
\subsection{$F(R,w)= R+\alpha w R $} \label{sec:FRW_alphawR}
In a specific case of $f(R)$, linear dependence of the Ricci scalar would be considered as: 
\begin{equation}\label{FRw_alpha}
F(R,w) = R+\alpha w R 
\end{equation}
The choice \eqref{FRw_alpha} appears to be a small perturbation to the theory for $\alpha\ll 1$. 
\\Interesting features can be found in this choice. First of all, The model coincides naturally with general relativity during the matter and radiation era, in which $w=0$ and $w=1/3$ respectively. Second, even with the non-minimal coupling with matter through the equation of state, the model appears as an effective energy-momentum theory with minimal coupling. Thus, the model is free from Einstein-Jordan frame ambiguity. 
\\ 
The field equations are given as: 
\begin{equation}
R_{\mu \nu}-\frac{1}{2}g_{\mu \nu} R=8\pi G T^{eff}_{\mu\nu}
\end{equation}
with 
\begin{equation}
T^{eff}_{\mu\nu}= \frac{ T_{\mu\nu}}{(1+\alpha w)}
\end{equation}
Friedmann equations can be shown as: 
\begin{equation}
 H^2= \frac{8\pi G}{3} \bigg(\frac{\rho^2}{\rho+\alpha p}\bigg)
\end{equation}
\begin{equation}
\dot{H}+H^2=-\frac{4\pi G}{3}\rho\bigg(\frac{\rho+3p}{\rho+\alpha p}\bigg)
\end{equation}
This class of models has great influence when considering scalar fields as the dominant matter in the universe to explain early or late inflationary expansions.
\subsection{$F(R,w)= f(R)+k(w) $}
As a third case of generalized $F(R,w)$ gravity models, we consider that the action is given by, 
\begin{equation}
F(R,w)= f(R)+k(w)
\end{equation}
For an arbitrary matter source the gravitational field equations are given by:
\begin{equation}
f_R R_{\mu \nu} -\frac{1}{2}g_{\mu\nu}f(R)=8\pi G (T_{\mu\nu}+K_{\mu\nu})
\end{equation}
where: 
\begin{equation}
K_{\mu\nu}=\frac{k(w)}{8\pi G}g_{\mu\nu}
\end{equation}
Once again, we can recover the minimal coupling case by applying the conformal transformation  $g_{\mu\nu}^E=\Phi g_{\mu\nu}^J $. 
\\
The Friedman equations take the form,
\begin{equation}
H_E^2=\frac{8\pi G}{3}(\rho_\Phi+\rho_M+\rho_w)_E
\end{equation}
\begin{equation}
(\dot{H}+H^2)_E=-\frac{4\pi G}{3}(2 \rho_\Phi-\rho_M+3(p_M+p_w))_E
\end{equation}
where $\rho_w=\frac{k(w)}{8\pi G}$, and $p_w=\frac{k(w)}{8\pi G}$
\section{Astrophysical implications:}
In order to understand the astrophysical implications of this gravitational extension, the function $F(R,w)$ needs to be defined specifically.
In this section, we examine the slow-roll inflationary scenario in the simplest case to demonstrate how $F(R,w)$ gravity impacts the main observable quantities represented by $n_s$ and $r$. 
Following that, we present three possible future research directions. Firstly, we provide a brief look at constant-roll inflation in which the $F(R,w)$ have a greater impact on the model. Next, we inspect the influence of $F(R,w)$ on Starobinsky's model, and finally we discuss the generalized polytropic equation of state. 
%\\In this section, howerver, we will foucus on the simplest case represented by equation [?]. Agian, the first intersting feature about this choice is recovering the minimalc coupling case withou the need to conformal transformation and the model is free of frame ambiguity. 
%\\
%In this sitiuation, we can find natural coincedence with gerenal relativity results in the matter dominated ear where $w=0$. For the the radiation dominated era where $w=1/3$, one can notice same coincedence with with general relativity for $\alpha \ll 1$. Howerver, considering significant values of the the parameter $\alpha$, can have a positive impact on the Neucleosynthesis in order to explain Y, in which general relativity faild to explain this. 
%However, this study will be performed in independent work. 
\subsection{Slow-roll Inflation:}
\begin{figure}
    \centering
    \includegraphics[width=1\linewidth]{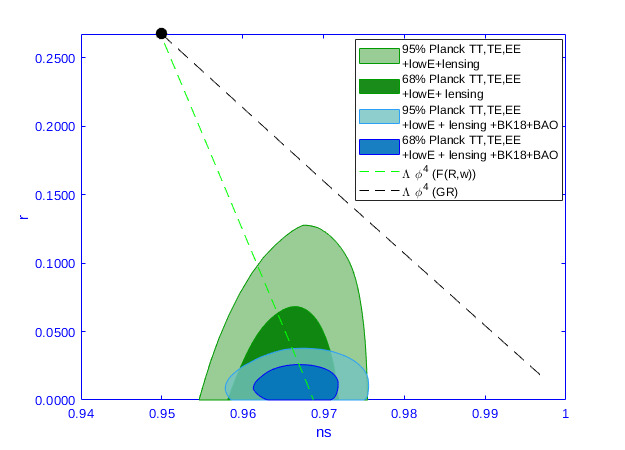}
    \caption{A plot of $n_s$ and $r$ in the case of the quartic potential . The black dashed line depicts the inflationary model within the framework of standard General Relativity (GR), while the green dashed line represents the model in F(R,w) gravity. The experimental contours are generated by combining the results of (Planck TT, TE, EE + lowE + lensing) denoted as "Green Contours," along with the results of (Planck TT, TE, EE + lowE + lensing + BK15) indicated as "Blue Contours" \cite{Planck:2018jri} .}
    \label{fig:eqwq}
\end{figure}

Using scalar fields as dominated matter, this section examines the early accelerated expansion of the universe. 
\\
Let's consider a canonical scalar field  as follows: 
\begin{equation}
L_\phi=-\frac{1}{2} g^ {\mu\nu}\nabla_\mu \phi \nabla_\nu \phi-V(\phi)
\end{equation}
with signature $(-,+,+,+)$.
\\ Following the results of section \eqref{sec:FRW_alphawR} and the function \eqref{FRw_alpha} , we find the following:
\begin{equation}\label{}
S=\int \sqrt{-g} d^4x \bigg[\frac{1}{2} R  + -\frac{1}{2} g^ {\mu\nu}\nabla_\mu \varphi \nabla_\nu \varphi -U(\phi(\varphi)) \bigg]
\end{equation}
Where the re-scaled canonical field $\varphi$ is given by: 
\begin{equation}
\frac{d\chi}{d\phi}=\frac{1}{\sqrt{(1+\alpha w)}}
\end{equation}
The potential $U$ is given by: 
\begin{equation}
U(\phi(\varphi)) = \frac{ V(\phi)}{(1+\alpha w)}
\end{equation}
As a result we can find Friedmann equation in Jordan frame as: 
\begin{equation}
H^2=\frac{1}{2}\dot{\chi}^2+U(\chi)
\end{equation}
\begin{equation}
\dot{H}=-\frac{1}{2}\dot{\chi}
\end{equation}
The equation of motion of the scalar field: 
\begin{equation}
\ddot{\chi}+3 H \dot{\chi}+U_\chi=0
\end{equation}
Let us now consider the most extreme case where the potential energy dominates the kinetic energy $U \gg X$, so $w\sim -1$. then the slow-roll approximation for constant $\alpha$ is given as: 
\begin{equation}
\varepsilon=\frac{(1-\alpha)}{2}\bigg(\frac{V_\phi}{V}\bigg)^2
\end{equation}
\begin{equation}
\eta=(1-\alpha)\bigg(\frac{V_{\phi\phi}}{V}\bigg)
\end{equation}
Thus, the spectral index $n_s$ and tensor to scalar ratio $r$ are represented as: 
\begin{equation}
n_s-1= ( 1-\alpha) ( 2\eta_\phi - 6 \varepsilon_\phi)
\end{equation}
\begin{equation}
r= 16 ( 1-\alpha) \varepsilon_\phi
\end{equation}
where $\varepsilon_\phi=\frac{1}{2}\bigg(\frac{V_\phi}{V}\bigg)^2$, and $\eta_\phi=\bigg(\frac{V_{\phi\phi}}{V}\bigg)$.
\\
Consider Higgs inflation with the potential given by: 
\begin{equation}
V(\phi)=\lambda (\phi^2-\nu)^2
\end{equation}
The potential takes the form of a quartic potential for $\nu\ll \phi$.
\\ Modified gravity models such as \cite{Pozdeeva:2020apf,Tahmasebzadeh:2016zhd,AlHallak:2023tca} take into account potentials of this type. In this section,However  we consider it in $F(R,w)$.
\\The observables $n_s$ and $r$ take the following forms: 
\begin{equation}
n_s= (1-\alpha) \bigg(1-\frac{24}{\phi^2}\bigg)
\end{equation}
\begin{equation}
r= (1-\alpha) \bigg(\frac{128}{\phi^2}\bigg)
\end{equation}
Although the value of $n_s$ and $r$ will be affected by the factor $(1-\alpha)$, the slope of the line representing the relationship between them remains unaffected ($\simeq 5.33$), so observational results are still exempt from the Higgs inflation, as we can see in Figure \eqref{fig:eqwq}.
\\
In this extremely simplified case of $w$, we can still see the benefits of the model by considering the parameter $\alpha(\phi)$. It is easy to see the effect by examining the term $(1-\alpha(\phi))$ as follows: 
\begin{equation}\label{eq:jfru}
(1-\alpha(\phi))=(1-\xi \phi^2)^2
\end{equation}
As can be seen in this case, the model is equivalent numerically to the non-minimal coupling case in Einstein frame \cite{Linde:2011nh}, but again, the treatment is performed in Jordan frame. 
\\
The non-minimal coupling to gravity has been extensively studied within different gravity formalisms, such as \cite{Linde:2011nh,Bauer:2008zj,AlHallak:2021sic,AlHallak:2016wsa,Karamitsos:2021mtb,Hertzberg:2010dc,Qiu:2010dk}. 
\\
Figure \eqref{fig:eqwq} showcases the theoretical results of the quartic potential within the context of the minimal coupling case in standard GR in contrast to the model within $F(R,w)$ gravity. The black dashed line represents the theoretical results obtained in the minimal coupling case. It is evident that the model is excluded since there is no intersection between the theoretical results and the observational contours. The black dot corresponds to the number of e-foldings $N=60$. The green dashed line corresponds to the results obtained in the $F(R,w)$ framework, taking into account the equation \eqref{eq:jfru}. The line corresponds to the value of N=60, with $\xi$ increasing from zero at the top to infinity at the bottom of the graph. The results of the inflationary model exhibit noteworthy agreement with the observational outcomes at the $95\% $and $68\%$ confidence levels of the combined results of (Planck TT, TE, EE + lowE + lensing), represented as "Green Contours," along with the results of (Planck TT, TE, EE + lowE + lensing + BK15), denoted as "Blue Contours."
\\ 
Keeping in mind that $\alpha$ can take different forms, one can expect and assess how various forms will impact inflationary results. In addition, we can consider the more natural case where $w=w(\phi)$, as in some inflationary scenarios, such as warm inflation, in which the inflaton field continuously dissipates into radiation during the inflationary period, or constant-roll inflation, where the kinetic term take place during the inflationary period. 
\subsection{Future Directions and Expected Implications}
In this section we provide some suggestions for future works to investigate important cosmological phenomena in the field of $F(R,w)$. We present here, some comments about this suggestions leaving the detailed investigations for future works. 
\subsubsection{Constant-roll inflation:}
The slow roll inflationary scenario was discussed in the previous section quickly to illustrate how $F(R,w)$ models of gravity impact cosmological inflation. 
\\
However, in the slow-roll inflationary scenario the kinetic energy term is negligible in front of the potential energy and the equation of state is approximately $-1$. As discussed earlier, $F(R,w)$ affects the inflationary model under the slow-roll approximation because of the coupling coefficient $\alpha$, which may be constant or depend on the scalar field. 

\begin{figure}
    \centering
    \includegraphics[width=1\linewidth]{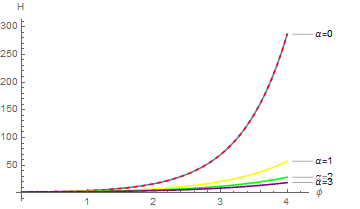}
    \caption{The Hubble parameter is plotted as a function of the scalar field for different $\alpha$ values in $F(R,w)$ gravity with $F(R,w)=R+\alpha R w$ compared to the GR model. A flatter Hubble parameter is seen as the parameter $\alpha$ increases.}
    \label{fig:erte}
\end{figure}

In this section, we discuss constant-roll inflation, in which kinetic energy has a significant role to play. 
\\Specifically, we consider $F(R,w)=k(w)R=(R+\alpha R w)$. 
\\
Accordingly, Friedmann equations will be expressed in terms of $\phi$ as follows: 
\begin{equation}\label{eq:fed}
3 H^2 = \frac{1}{k(w)}[ \frac{1}{2}\dot{\phi}^2+V(\phi) ]
\end{equation}
\begin{equation}\label{eq:dotH}
- 2 \dot{H}=\frac{1}{k(w)} \dot{\phi}^2
\end{equation}
The scalar field's equation of motion is: 
\begin{equation}
\ddot{\phi}+3H\dot{\phi}+k(w) U_\phi=0
\end{equation}
Where $U_\phi$ represents the derivative of the potential $U$ with respect to scalar field $\phi$. A minimum ground state $\dot{w}=0$ is assumed in the previous equations. 
\\
As part of the original work of the constant-roll inflation \cite{Motohashi:2014ppa}, authors used the Hamiltonian-Jacobi formalism to extract the Hubble parameter from the inflaton field $H = H(\phi)$. 
\\
In this paper we follow same argument which is applicable as long as $t = t(\phi)$ is a single-valued function, i.e $\dot{\phi} \neq 0$.
\\
The constant rate of rolling is parameterized by the equation: 
\begin{equation}\label{eq:const_rat_condition}
\ddot{\phi}=-(3+\beta)H \dot{\phi}
\end{equation}
As shown in ref \cite{Motohashi:2014ppa}, $H=H(\phi)$ is independent of the inflationary model represented by $V(\phi)$, and its general solution can be formulated as follows: 
\begin{equation}
H(\phi)=C_1 \exp\bigg(\sqrt{\frac{3+\beta}{2}}\frac{\phi}{M_{pl}}\bigg)+C_2 \exp\bigg(-\sqrt{\frac{3+\beta}{2}}\frac{\phi}{M_{pl}}\bigg)
\end{equation}
\\
Conversely, we can find in the $F(R,w)$ from equation \eqref{eq:dotH} that: 
\begin{equation}\label{fsdr}
- 2 M_{pl}^2 H_\phi = \frac{1}{k(w)} \dot{\phi}
\end{equation}
We can determine $\ddot{\phi}$ from the time derivatives of equation \eqref{fsdr}. 
\\The differential equation for the Hubble parameter $H$ can be found by replacing $\ddot{\phi}$ in equation \eqref{eq:const_rat_condition} as, 
\begin{equation}\label{eq:kfut}
H_{\phi\phi}=\frac{(3+\beta)}{2M_{pl^2}k(w)}H
\end{equation}
 Solving equation \eqref{eq:fed} , one can find $\dot{\phi}(\phi)$ as, 
\begin{equation}
\dot{\phi}^2=3\alpha M_{pl}^2 H^2+\sqrt{3\alpha M_{pl}^2 H^2(3\alpha M_{pl}^2 H^2-8 V)-2V}
\end{equation}
In the case of quartic potential $ V=\lambda \phi^4$, we solved the differential equation \eqref{eq:kfut} numerically.
\\ In figure \eqref{fig:erte}, we plot $H$ as a function of $\phi$ in both standard GR \cite{Motohashi:2014ppa} and $F(R,w)$ for comparison.
\\ As shown in figure \eqref{fig:erte}, the Hubble parameter's function in GR and $F(R,w)$ are exactly the same for $\alpha=0$, as it expected theoretically. 
\\
Here is a numerical example where each parameter value of the model is fixed to unity, so that $H$ is represented in figure \eqref{fig:erte}. Given $\phi=4$, we found $\frac{H_{F(R,w)}}{H_{\text{GR}}}=0.2$ for $\alpha=1$. The ratio $\frac{H_{F(R,w)}}{H_{\text{GR}}}$ depends on $\alpha$, where for $\alpha=2$ results in a ratio $0.1$, while $\alpha=3$ results in a ratio $0.07$. 
\\
 As a result, we can conclude that the model $F(R,w)$ has an impact on the evolution of the universe. 
 \\On the other hand, when checking the effect of the energy scale $\lambda$ of the potential $V$, we found that there is no effect at all. Accordingly, the question arises as to whether the constant-roll inflation in $F(R,w)$ gravity is independent of the potential, similar to general Relativity, or not. This conclusion, however, requires more investigation, which can be carried out in future studies. All we are looking at here is the estimated inflationary impact of the $F(R,w)$ model. 
\subsubsection{Starobinsky cosmological model: }
\begin{figure}
    \centering
    \includegraphics[width=1\linewidth]{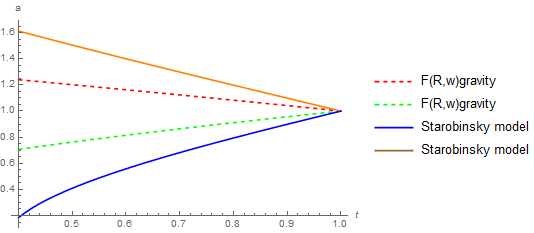}
    \caption{The scale factor for $F(R,w)=f(R)(1+\alpha w)$ is represented by dashed lines compared with the pure Starobinsky model, which represents $f(R)=R+\gamma R^2$  by solid lines.}
    \label{fig:ewq}
\end{figure}

\begin{figure}
    \centering
    \includegraphics[width=1\linewidth]{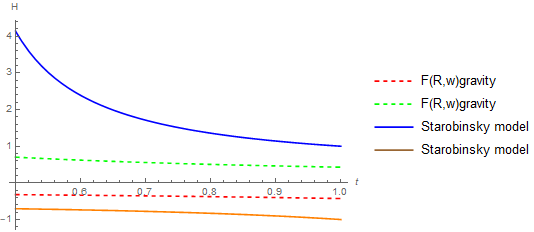}
    \caption{The Hubble parameter for $F(R,w)=f(R)(1+\alpha w)$ is represented by dashed lines in contrast to the pure Starobinsky model, which represents $f(R)=R+\gamma R^2$  by solid lines.}
    \label{fig:few}
\end{figure}

Assuming the function $F(R,T)=f(R)k(w)=(R+\gamma R^n)(1+\alpha w)$, then the trace of equation \eqref{eq:erh} is shown as: 
\begin{equation}\label{fdsf}
R=\gamma(n-2)R^n=-T^{\text{eff}}
\end{equation}
For $\gamma \gg 1$ and $n\neq 1$ we can write \eqref{fdsf} as:
\begin{equation}\label{fswe}
R=\bigg(-\frac{T^{\text{eff}}}{\gamma (n-2)}\bigg)^\frac{1}{n}
\end{equation}
In the Starobinsky model, where $f(R)=R+\gamma R^2$, the Ricci scalar is:   
\begin{equation}\label{rwe}
R=-T^{\text{eff}}=\frac{(3P-\rho)\rho}{\alpha P}
\end{equation}
To simplify the analytical equation, we considered $\alpha w \gg 1$.
\\Considering $\Lambda$CDM model with flat FRW, we can derive Fiendmann equation as follows: 
\begin{equation}\label{fsdfe}
\begin{split}
H^2=\big(\frac{b^2}{b+d/2}\big)^2 \frac{3(\rho_m+\rho_r+\rho_\Lambda)}{\alpha(\rho_r-3\rho_\Lambda)} \\ \bigg[\frac{3(\rho_m+\rho_r+\rho_\Lambda)}{\alpha(\rho_r-3+\rho_\Lambda)}(\rho_m+\rho_\Lambda)^2
\\ 
\frac{(K-3)(K+1)}{2b}+(\rho_m+\rho_\Lambda)+\rho_r/b\bigg]  
\end{split} 
\end{equation}
where: 
\begin{equation}\label{rerw}
K=\frac{3 \rho_\Lambda}{(\rho_m+\rho_\Lambda)}
\end{equation}
\begin{equation}\label{r34}
b=1+\frac{6 \gamma}{\alpha} (\rho_m+4\rho_\Lambda) \frac{(\rho_m+\rho_r+\rho_\Lambda)}{(\rho_r-3\rho_\Lambda)}
\end{equation}
\begin{equation}\label{4234}
d=\frac{\dot{b}}{H}
\end{equation}
A numerical solution of equation \eqref{fsdfe} is used to determine the factors of universe expansion represented by scale factor and Hubble parameter. These results are presented along with the results within the Starobinsky model in figures \eqref{fig:ewq} and \eqref{fig:few}. 
\\
The numerical check proceeds by fixing unit values for all parameters except the parameter $\beta=4$, and assuming current value of scale factor equals to one. Figure \eqref{fig:ewq} shows two solutions for the scale factor in $F(R,w)$ "dashed lines", of which the "Brown curve" corresponds to a negative Hubble's parameter solution, as seen in figure \eqref{fig:few}. Starobinsky's pure $F(R)$ model also has similar solutions, although with different amplitudes. 
\\A significant impact is expected on the grand scheme of evolution of the universe as a whole as a result of $F(R,w)$. Nevertheless, a detailed study should be carried out in the future.  
%\begin{figure}
%    \centering
%    \includegraphics[width=0.75\linewidth]{adotdot.png}
%    \caption{Enter Caption}
%    \label{fig:432}
%\end{figure}
\subsubsection{Polytropic equation of state:}
All of the last discussions focused on the perfect fluid's equation of state as $P=\kappa \rho$, where $\kappa=0$ represents dust matters, $\kappa=1/3$ corresponds to radiation, and $\kappa=0$ represents vacuum energy. In order to understand different cosmological aspects, different equations of state may be required, such as a polytropic equation of state in which $P = \xi \rho^\sigma$.
where $\xi$ and $\sigma$ are constants.$\sigma$ usually is written as, $\sigma = 1- 1/n$, where $n$ is known as polytropic index.
 \\ Reference \cite{Mardan:2020noh}  shows how with a generalized polytropic equation of state we can obtain new classes of polytropic models from the solution of Einstein-Maxwell field equations for charged anisotropic fluid configurations. Various values of polytropic index $n = 1,2$ are used to develop the models. Models that have been developed have helped regain masses and radii of eight different stars. Analyses of the models indicate that they are well-behaved and physically feasible.

In  \cite{Chavanis:2021jwr} authors develop a cosmological model involving a scalar field with a power-law potential associated with polytropic equations of state. In order to simplify the equations of the problem, they consider a fast oscillation regime called "spintessence". Various values of polytropic constants and polytropic indexes are studied. The LambdaCDM model, the Chaplygin gas model, and the Bose-Einstein condensate model are recovered.
\\
Considering such an equation of state and examining models in F(R,w) are worthy of investigation, and a significant impact can be expected.
\section{Summary and Conclusion:}
A modified gravitational model is proposed in this paper. The modification is represented by the function $F(R,w)$, where $w$ is the equation of state. 
\\
Three different cases are discussed with functions $F(R,w) = f(R) k(w)$,$F(R,w) =f(R)+k(w)$ and $F(R,w) =(R+\alpha R w)$. The models are studied under the Palatini formalism.
\\ The first and second choices appear as $F(R)$ models with an effective momentum-energy tensor. In contrast, the third choice takes the form of a Hilbert-Einstein action with an effective momentum-energy tensor. The third case shows no Jordan-Einstein frame ambiguity.  
\\
The slow-roll inflation model is examined to illustrate how $F(R,w)$ may appear to mimic the standard non-minimal coupling to gravity.
\\
We consider the Higgs inflationary potential as a case study and demonstrate how the $F(R,w)$ model effectively addresses the challenges posed by Higgs inflation.
\\
Finally, we present three future directions for work and provide some initial indications of how $F(R,w)$ models could impact cosmological results.
\\ A constant-roll inflation model with $ w=w(\phi)$ is discussed, and the Hubble parameter is numerically impacted by this model. 
\\ In the $\Lambda$ CDM model, specifically under the $F(R,w)=F(R) (1+\alpha w)$ framework with $F(R)=R+\gamma R^2$, we examine the behavior of the scale factor and Hubble parameter using the Starobinsky function. We compare the results to a pure Starobinsky model to clarify the expected effect of $F(R,w)$.
\\As a final point, we introduce the Polytropic equation of state as a field that will be of interest to future works.

\end{document}